\documentclass[12pt,thmsa]{article}
\usepackage{amssymb}

\usepackage{sw20lart}

\input{tcilatex}
\begin{document}

\smallskip {\large Universal Construction of Unitary Transformation of }%
\newline
{\large Quantum\ Computation with One- and Two-body \smallskip Interactions}

{\normalsize \smallskip }

\smallskip

\medskip

\ \ \quad \qquad \qquad \qquad \qquad Xijia Miao* \newline
Laboratory of Magnetic Resonance and Atomic and Molecular Physics, Wuhan
Institute of Physics and Mathematics, The Chinese Academy of Sciences, Wuhan
430071, P.R.China; Department of Biochemistry, The Hong Kong University of
Science and Technology, Clear Water Bay, Kowloon, Hong Kong \newline
\newline
*correspondence address: Wuhan Institute of Physics and Mathematics. E-mail:
miao@nmr.whcnc.ac.cn \newline
April 16, 2000 \newline
\newline
\textbf{Abstract}

\smallskip Any unitary transformation of quantum computational networks is
explicitly decomposed, in an exact and unified form, into a sequence of a
limited number of one-qubit quantum gates and the two-qubit diagonal gates
that have diagonal unitary representation in usual computational basis. This
decomposition may be simplified \smallskip greatly with the help of the
properties of the finite-dimensional multiple-quantum operator algebra
spaces of a quantum system and the specific properties of a given quantum
algorithm. As elementary building blocks of quantum computation, the
two-qubit diagonal gates and one-qubit gates may be constructed physically
with one- and two-body interactions in a two-state quantum system and hence
could be conveniently realized experimentally. The present work will be
helpful for implementing generally quantum computations with any qubits in
those feasible quantum systems and determining conveniently the time
evolution of these systems in course of quantum computation. \newline
\newline
\textbf{1. Introduction }

Since it has been discovered that quantum computers can be much more
powerful than their classical counterparts (Feynman 1982; Deutsch 1985; Shor
1994), it becomes of great practical importance to realize the quantum
computers. A variety of quantum systems have been explored to build such
quantum computers such as trapped ions (Cirac \& Zoller 1995), nuclear spins
in molecules (Gershenfeld \& Chuang 1997; Cory et al. 1997) and in solid
states (kane 1998), and Josephson junction arrays in superconductors
(Makhlin et al. 1999), etc. Very recently, nuclear magnetic resonance (NMR)
(Ernst et al. 1987; Freeman 1997) is used to realize experimentally the
Deutsch-Jozsa$^{^{\prime }}$s algorithm (Deutsch \& Jozsa 1992, Chuang et
al. 1998a) and the Grover$^{^{\prime }}$s algorithm (Grover 1997, Chuang et
al. 1998b, Jones et al. 1998). It is expected that in the short term quantum
computation with as many as ten qubits will be implemented, although there
are many problems that need to be solved such as how to overcome the effects
of decoherence and dephase in quantum systems and how to realize
fault-tolerant and error-correlating quantum computation. Quantum
computation should be performed within the characteristic time of
decoherence and dephase in a quantum system. In practice, this requires that
elementary building blocks of quantum computation should be chosen suitably.
For example, theoretically any quantum computation should be built exactly
out of a sequence of the building blocks with a length as short as possible,
while these building blocks can be exactly constructed theoretically and
could be physically realized conveniently in a feasible quantum system.
Quantum gates were firstly suggested by Deutsch (1989) as the elementary
building blocks to construct any quantum computational networks. A universal
quantum gate, by copying itself and then wiring together, suffices to
construct any unitary transformation of quantum computation. It has been
shown theoretically (Deutsch et al. 1995, Lloyd 1995) that almost every
quantum gate that operates on two or more qubits is universal, but the
universal quantum gates that are really considered as conveniently
realizable gates are the three-qubit gates, e.g., Toffoli$^{^{\prime }}$s
(1981), Fredkin$^{^{\prime }}$s (1982), and more general Deutsch$^{^{\prime
}}$s (1989) gate and the simpler two-qubit gates (Barenco 1995 \& Sleator
and Weinfurter 1995). Quantum gates with more qubits may not be universally
attractive since their construction and implementation may be usually
complicated in quantum systems and the theoretical construction of any
unitary transformation of quantum computation out of these gates is usually
not carried out easily. This may be the main reason why many investigators
(DiVincenzo 1995a; Barenco 1995 \& Sleator and Weinfurter 1995) have
suggested the simpler universal quantum gates with only two qubits as the
building blocks. Barenco et al. showed further that a set of quantum gates
that consists of all one-qubit gates and the two-qubit XOR gate suffices to
build any unitary transformation, although one-qubit gates and the simpler
two-qubit XOR gate are not universal. However, the present theoretical
composition of the three- or two-qubit universal gates or even the XOR gate
along with one-qubit gates to form quantum networks usually disregards some
useful properties of quantum systems and the specific properties of a given
quantum computation. As a consequence, it needs probably an infinite number
of such quantum gates to build exactly a given unitary transformation of
quantum computation. Obviously, this is impractical for quantum computation
to be performed in a quantum system in which the effects of decoherence and
dephase are not negligible.

In this paper any unitary transformation of quantum computational networks
is decomposed explicitly as a sequence of a limited number of one- and
two-body elementary propagators, i.e., one-qubit gates and the two-qubit
diagonal gates $R_{kl}(\lambda _{kl})=\exp (-i\lambda _{kl}2I_{kz}I_{lz})$
that have diagonal unitary representation in the conventional computational
basis, in an exact and unified form. Quantum computation with any qubits is
then implemented by performing the unitary transformation of a sequence of
these one- and two-qubit gates on the input quantum state. In constrast to
the usual composition of quantum gates to form quantum networks here is
emphasized on the decomposition of a given quantum network as a sequence of
the elementary building blocks. This decomposition may be simplified greatly
with the aid of the properties of the finite-dimensional multiple-quantum
operator algebra spaces of quantum systems (Miao, 2000a) and the specific
properties of a given quantum algorithm. That the two-qubit diagonal gate $%
R_{kl}(\lambda _{kl})$ is chosen as an elementary building block stems from
several considerations. Firstly, such choice for building blocks is
beneficial to the exploitation of the properties of the finite-dimensional
multiple-quantum operator algebra spaces of quantum systems to simplify the
decomposition. Secondly, the two-qubit diagonal gates supplemented with all
one-qubit gates suffice to construct exactly any unitary transformation of
quantum computation. Thirdly, from the physically realizable point of view
one-qubit gates should be the simplest gates, while according to matrix
(operator) algebra properties quantum gates that have diagonal unitary
representation matrices in the conventional computational bases $%
\{|00...00\rangle ,|00...01\rangle ,...,|11...11\rangle \}$ are the
elementary and simple gates. Particularly, the two-qubit diagonal gate $%
R_{kl}(\lambda _{kl})$ that has diagonal unitary representation matrix:

$\qquad \qquad \qquad R_{kl}(\lambda _{kl})=\left[ 
\begin{array}{llll}
e^{-i\frac{1}{2}\lambda _{kl}} &  &  &  \\ 
& e^{i\frac{1}{2}\lambda _{kl}} &  &  \\ 
&  & e^{i\frac{1}{2}\lambda _{kl}} &  \\ 
&  &  & e^{-i\frac{1}{2}\lambda _{kl}}
\end{array}
\right] $\newline
\newline
should be the most elementary and simplest building blocks, although it is
not a universal gate. Moreover, the two-qubit diagonal gate $R_{kl}(\lambda
_{kl})$ can be conveniently built up with one- and two-body interactions
such as the general neighbor interaction and could be easily realized in a
quantum system with N two-state particles such as coupled multispin systems
in molecules or in solid states, trapped ions, superconducting Josephson
junction arrays, etc. Fourthly, the two-, three-, and N-qubit (N\TEXTsymbol{>%
}3) universal gates and even the XOR gate can be expressed exactly as a
simple sequence of the two-qubit diagonal gates along with one-qubit gates.%
\newline
\newline
\textbf{2. The decomposition of unitary transformation}

\smallskip Quantum computation is a reversible process (Bennett, 1973). It
can be thought of as a unitary transformation acted on the input state and
obeys the laws of quantum mechanics (Benioff, 1980; Deutsch, 1985;
DiVincenzo, 1995b). The time evolution of a quantum system from the initial
state to the output state during quantum computation then can be described
by a time-evolutional propagator $U(t)$ that obeys the Schr$\ddot{o}$dinger
equation:

{\large \ }$\qquad \dfrac{d}{dt}U(t)=-iH(t)U(t)\qquad (\hslash =1)\qquad
\qquad \qquad \qquad \qquad \ \ \ \ \qquad \ (1)\newline
$where $H(t)$ is the effective Hamiltonian for the system to perform the
quantum computation. The effective Hamiltonian $H(t)$ characterizes
generally the specific properties of the quantum computation. In general,
quantum computational networks are composed of a sequence of quantum circuit
units (Deutsch, 1989). Each such circuit unit performs the unitary
transformation with a propagator $U_{k}(t_{k})$ associated with the
time-independent effective Hamiltonian $H_{k}$ in the interval $t_{k}$,
while the total propagator $U(t)$ can be expressed as a sequence of the
propagators $U_{k}(t_{k})$. Then it follows from Eq.(1) that

$\qquad U(t)=\stackunder{k}{\prod }U_{k}(t_{k})=\stackunder{k}{\prod }\exp
(-iH_{k}t_{k})\qquad (t=\stackunder{k}{\sum }t_{k})$ $\qquad \qquad \qquad
\quad \ (2)$ \newline
The quantum computation then can be implemented by acting a sequence of the
propagators $U_{k}(t_{k})$ on the input state. Therefore, it becomes clear
that the problem to be solved is how to exactly decompose theoretically the
propagators $U_{k}(t_{k})$ of quantum circuit units as a sequence of a
limited number of one- and two-qubit gates and how to build up these simple
gates experimentally in an accessible quantum system. Not loss of
generality, the quantum system is considered as a physical system consisting
of N two-state particles. This system may be nuclear spins in molecules or
in solid state, trapped ions, and superconducting Josephson junctions, etc.
Here for simplification the complete decomposition of the propagators is
described explicitly in a coupled spin (I=1/2) system, which is formed by N
two-state nuclei with magnetic quantum number I=1/2.

The effective Hamiltonian $H_{j}$ associated with each quantum circuit unit
can be generally expanded as a linear combination of base operators $%
\{B_{k}\}$ of the Liouville operator space of the spin system (Ernst, et al.
1987):

$\qquad \qquad \qquad \qquad \qquad \ \ H_{j}=\stackunder{k}{\sum }%
a_{k}B_{k}\qquad \qquad \qquad \qquad \qquad \qquad \qquad (3)$ \newline
As suggested recently (Miao, et al. 1993 \& 1997; Miao, 2000a), to determine
exactly and analytically time evolution of the spin system the propagator
corresponding to this Hamiltonian is first decomposed into an ordered
product of a series of elementary propagators

$\qquad \qquad \qquad \qquad U_{j}(t_{j})=\exp (-iH_{j}t_{j})=\stackunder{s}{%
\prod }R_{s}(\lambda _{s})\qquad \qquad \qquad \qquad (4)$\newline
The elementary propagator is defined by

$\qquad \qquad \qquad \qquad R_{s}(\lambda _{s})=\exp (-i\lambda
_{s}B_{s})\qquad \qquad \qquad \qquad \qquad \ \ \ \qquad (5)$ \newline
where $\lambda _{s}$ is a real parameter and $B_{s}$ a Hermite base
operator. Obviously, the elementary propagator is also a quantum gate.
Actually, the decomposition of Eq.(4) can be achieved in an exact and
unified form. Firstly, the propagator $U_{j}(t_{j})$ is converted unitarily
into a diagonal unitary operator, which has diagonal unitary representation
in usual computational basis, by making a sequence of elementary unitary
transformations. Then each such elementary unitary transformation and the
diagonal unitary operator are further decomposed into a product of a series
of elementary propagators, respectively. The decomposition of Eq.(4) can be
further simplified with the help of the properties of the Liouville operator
spaces and its three subspaces (Miao, 2000a): the even-order
multiple-quantum, the zero-quantum, and the longitudinal magnetization and
spin order operator subspace. When the effective Hamiltonian $H_{j}$ is a
member of the longitudinal magnetization and spin order operator subspace,
the propagator $U_{j}(t_{j})$ is simply expressed as a sequence of
elementary propagators built up with the base operators of the subspace (see
below). If $H_{j}$ is a member of the zero-quantum operator subspace, one
first makes a zero-quantum unitary transformation on $U_{j}(t_{j})$ to
convert it into the diagonal unitary operator and then further decomposes
the zero-quantum unitary operator and the diagonal unitary operator as a
sequence of elementary propagators, respectively. When the effective
Hamiltonian $H_{j}$ is a member of the even-order multiple-quantum operator
subspace, one makes the even-order multiple-quantum and subsequently the
zero-quantum unitary transformation on $U_{j}(t_{j})$ to convert it into the
diagonal unitary operator. The even-order multiple-quantum, the
zero-quantum, and the diagonal unitary operator can be further decomposed as
a sequence of elementary propagators, respectively. If $H_{j}$ is not a
member of any one of the above three subspaces but a member of the Liouville
operator space, one first converts it unitarily into a member of the
even-order multiple-quantum operator subspace by making an odd-order
multiple-quantum unitary transformation on the Hamiltonian, then a further
decomposition for the propagator $U_{j}(t_{j})$ can be carried out with the
help of the properties of the even-order multiple-quantum operator subspace.

It is clearly shown from the closed property of operator algebra space that
any quantum gate built up with an arbitrary operator of any one of the three
aforementioned operator subspaces is a non-universal gate. These gates can
form another set of non-universal gates that may be different from one-qubit
gates and collection of one-qubit gates and the classical gates (Deutsch,
1995).

On the basis of the decomposition of Eq.(4) time evolution of a system in
the course of quantum computation can be determined directly by acting the
decomposed propagator on the input state in a quantum system or on the
initial density operator in a quantum ensemble with the help of the rotation
transformation between any two base operators. The rotation transformation
can be generally derived from the Baker-Campbell-Hausdoff formula (Ernst, et
al. 1987):

$\qquad \qquad \qquad \qquad R_{s}(\lambda _{s})B_{r}R_{s}(\lambda
_{s})^{-1}=\stackunder{n=0}{\sum }\frac{(-i\lambda _{s})^{n}}{n!}C_{n}\qquad
\qquad \qquad \qquad (6)$ \newline
where $C_{0}=B_{r}$ and $C_{n}=[B_{s},C_{n-1}]$ and particularly, if $%
[B_{s},[B_{s},B_{r}]]=\alpha B_{r}$ the transforamtion (6) reduces to a
simpler closed form

$\qquad \ R_{s}(\lambda _{s})B_{r}R_{s}(\lambda _{s})^{-1}=B_{r}\cos (\sqrt{%
\alpha }\lambda _{s})-\frac{i}{\sqrt{\alpha }}[B_{s},B_{r}]\sin (\sqrt{%
\alpha }\lambda _{s})\qquad \ \ (7)$

For a coupled N-spin (I=1/2) system the proper base operators $\left\{
B_{k}\right\} $ of the Liouville operator space are usually chosen as the
Cartesian product operators (S$\phi $rensen, et al. 1983; Ernst, et al.
1987):

$\qquad \left\{ B_{k}\right\} =\{E,I_{k_{1}\alpha },2I_{k_{1}\alpha
}I_{k_{2}\beta },...,2^{n-1}I_{k_{1}\alpha }I_{k_{2}\beta }...I_{k_{n}\delta
};1\leq n\leq N\}\qquad (8)$ \newline
where $E$ is unit operator and $I_{k_{i}\alpha }$ $(\alpha ,\beta ,\delta
=x,y,z)$ are spin angular momentum operators for the $k_{i}$th spin in the
system ($I_{k_{i}}=\frac{1}{2}\sigma _{k_{i}},\sigma $ is the Pauli$%
^{^{\prime }}$s operator). Such direct product operator set contains any
n-body $(N\geq n\geq 1)$ interaction terms $\{2^{n-1}I_{k\alpha }I_{l\beta
}...I_{m\delta }\}$. It follows from Eqs.(2), (4), and (5) that the
propagator $U_{j}(t_{j})$ is usually expressed as a sequence of elementary
propagators built up with any n-body $(N\geq n\geq 1)$ product operators in
set (8). Actually, the propagator $U_{j}(t_{j})$ can be further expressed as
a sequence of the elementary propagators built up only with one- and
two-body operators of the set (8).

Each base operator of set (8) can be converted unitarily into a member of
the longitudinal magnetization and spin order operator subspace (Miao,
2000a), where the base operators of the subspace are usually chosen as the
longitudinal magnetization and spin order product operators in the N-spin
(I=1/2) system (Miao, et al. 1993; Miao, 2000a):

$\qquad
\{B_{k}^{m}\}=%
\{E,I_{kz},2I_{kz}I_{lz},4I_{kz}I_{lz}I_{mz},...,2^{N-1}I_{1z}I_{2z}...I_{Nz}\}\qquad \ \ \qquad (9) 
$ \newline
Then any elementary propagator defined by Eq.(5) can be converted unitarily
into an elementary propagator built up with a product operator of the
subspace by applying a limited number of 90 degree electromagnetic pulses. A
typical example is shown below

$I_{kx}I_{ly}...I_{my}\dfrac{\exp (i\frac{\pi }{2}I_{ky})} {}\dfrac{\exp (-i%
\frac{\pi }{2}I_{lx})} {}......\dfrac{\exp (-i\frac{\pi }{2}I_{mx})} {}\
I_{kz}I_{lz}...I_{mz}$ \newline
where the unitary transformation $B=UAU^{+}$\smallskip is denoted briefly as 
$A\frac{\ \ U\ \ } {}\ B$. Therefore, the basic building blocks for the
propagator $U_{j}(t_{j})$ are those elementary propagators built up with the
base operators of the subspace and the one-body elementary propagators of
Eq.(5). On the other hand, any elementary propagator constructed with an
n-body $(N\geq n\geq 1)$ product operator of the subspace can be readily
decomposed as a product of a series of the elementary propagators built up
only with the two-body product operators in set (9) and the one-body base
operators in set (8). This can be achieved by utilizing recurrently the
following decomposition:

$\exp (-i\lambda 2^{n}I_{k_{1}z}...I_{k_{n}z}I_{k_{n+1}z})=$

$\qquad \qquad \quad V_{n}\exp (-i\lambda
2^{n-1}I_{k_{1}z}...I_{k_{n-1}z}I_{k_{n+1}z})V_{n}^{+}$ $\quad (n\geq
2)\qquad \ \ \qquad (10)\newline
$where

$V_{n}=\exp (-i\frac{\pi }{2}I_{k_{n+1}x})\exp (-i\pi
I_{k_{n}z}I_{k_{n+1}z})\exp (i\frac{\pi }{2}I_{k_{n+1}x})\exp (-i\frac{\pi }{%
2}I_{k_{n+1}y})\qquad $\newline
As a consequence, it follows from Eqs.(4) and (5) that the propagators $%
U_{j}(t_{j})$ of the quantum circuit units and hence the total propagator $%
U(t)$ of quantum computation can be decomposed completely into a product of
a series of one-body elementary propagators and the two-body diagonal
elementary propagators $R_{kl}(\lambda _{kl})$ built up with the product
operators $\{2I_{kz}I_{lz}\}$.

Evidently, any operator of the longitudinal magnetization and spin order
operator subspace has the diagonal representation in usual computational
basis and any two base operators $B_{r}^{m}$ and $B_{s}^{m}$ of the subspace
are commutable with each other. Then any diagonal operator of the system,
i.e., an operator that has diagonal representation in usual computational
basis, can be expressed as a sum of the base operators of the subspace
(Miao, 2000a). If the effective Hamiltonian $H_{j}\ (=\stackunder{k}{\sum }%
a_{k}B_{k}^{m})$ associated with a quantum circuit unit is a member of the
subspace, the corresponding propagator $\exp (-iH_{j}t_{j})$ can be readily
decomposed as a product of a series of elementary propagators constructed
with the base operators of the subspace:

$\qquad \qquad \qquad \qquad \exp (-iH_{j}t_{j})=\stackunder{k}{\prod }\exp
(-ia_{k}B_{k}^{m}t_{j})\qquad \quad \ \ \qquad \qquad (11)$ \newline
Equation (11) is very useful for the decomposition of the total propagator
of a given quantum algorithm.\newline
\newline
\textbf{3. Preparation of the elementary building blocks }\smallskip

The two-qubit diagonal quantum gate $R_{kl}(\lambda _{kl})$ could be easily
prepared in many two-state physical systems. As an example, its preparation
is described explicitly in an accessible coupled N-spin (I=1/2) system. In
general, the external electromagnetic field such as radiofrequency (RF)
field is used to control the process of quantum computation and in the
coupled spin (I=1/2) system with ravelled resonances each one-body
elementary propagator defined by Eq.(5) may be prepared by utilizing
selective pulses (the weak RF field) (Freeman, 1997). The spin Hamiltonian
for the system in a strong static magnetic field is written as (Ernst, et
al. 1987)

$\qquad \qquad \qquad \qquad H_{0}=\stackunder{k}{\sum }\Omega _{k}I_{kz}+%
\stackunder{k<l}{\sum }\pi J_{kl}2I_{kz}I_{lz}\qquad \qquad \qquad \qquad \
\ (12)$ \newline
where it is assumed that the internuclear interaction is weak with respect
to the Zeeman interaction and the interaction between the system and its
environment, which results in decoherence and dephase, is negligible. This
Hamiltonian that consists of one-body $\left\{ I_{kz}\right\} $ and two-body 
$\left\{ 2I_{kz}I_{lz}\right\} $ interactions is responsible for preparing
experimentally the two-qubit diagonal quantum gates $R_{kl}(\lambda _{kl})$.
Figure 1 presents the quantum circuit unit (the NMR\ pulse sequence) for the
preparation of the elementary propagator built up with the direct two-body
interaction between two spins $k$ and $l$, where spin echoes refocus all the
undesired one- and two-body interactions and only leave selectively the
desired two-body interaction $2I_{kz}I_{lz}$ in the Hamiltonian (12) by
combining selective 180 degree pulses. If there is not direct interaction
between any two spins $k$ and $m$, their indirect two-body interaction $%
2I_{kz}I_{mz}$ may be achieved through a directly neighbor coupling network
such as $k-l-...-s-t-m$ in the system:

$2I_{kz}I_{mz}\dfrac{\exp (-i\pi I_{kx}I_{rx})} {}\dfrac{\exp (-i\pi
I_{ky}I_{ry})} {}\ 2I_{rz}I_{mz}\dfrac{\quad } {}......\dfrac{\quad } {}\
2I_{sz}I_{mz}$

$\qquad \qquad \dfrac{\exp (-i\pi I_{sx}I_{tx})} {}\dfrac{\exp (-i\pi
I_{sy}I_{ty})} {}2I_{tz}I_{mz}$\newline
For quantum dots (Barenco, et al. 1995a; Loss \& DiVincenzo, 1998) the
diagonal gates $R_{kl}(\lambda _{kl})$ could be prepared in an analogous way
to the above approach. In trapped ion system (Cirac \& Zoller, 1995) the
diagonal gates could be implemented by six laser pulses (see Appendix B) and
in superconducting Josephson junction arrays (Makhlin, et al. 1999) they
could also be prepared easily (Miao, 2000b). \newline
\newline
\textbf{4. Application to the universal quantum gates and quantum algorithms}

It is easy to carry out the explicit decomposition of the total propagators
for N-qubit quantum algorithms such as the Deutsch-Jozsa, Grover, quantum
Fourier transform algorithm, etc. and for the two-, three-, and N-qubit (N%
\TEXTsymbol{>}3) universal quantum gates. Several typical examples are given
explicitly below.\newline
\newline
\textbf{4.1\ The two-, three-, and any N-qubit universal quantum gates }

\smallskip The unitary representation matrix $U_{N}$ of the N-qubit
universal gate ( Deutsch, 1989; Barenco, 1995 \& Barenco, et al. 1995b) can
be generally written as

$U_{N}=E+Diag(0,0,...,0,1)\bigotimes (\left[ 
\begin{array}{ll}
-1 & \ \ 0 \\ 
\ \ 0 & -1
\end{array}
\right] +\left[ 
\begin{array}{ll}
u_{11} & u_{12} \\ 
u_{21} & u_{22}
\end{array}
\right] )\qquad \qquad \ (13)$ \newline
where the matrix $[u_{ij}]$ acting on the $N$th qubit is $u(2)$ unitary
matrix and can be generally expressed as

$\qquad \qquad \qquad U(2)=T_{N}\exp [-i(\varphi _{0}+\varphi
_{1}I_{Nz})]T_{N}^{+}\qquad \qquad \quad \qquad \qquad (14)$\newline
where $T_{N}=\exp (-i\alpha I_{Nz})\exp (-i\beta I_{Ny})$. Then the unitary
operation $U_{N}$ can be decomposed completely as a simple sequence of
one-body elementary propagators and the two-body diagonal elementary
propagators:

$\qquad \qquad \qquad U_{N}=T_{N}\exp (-i\widetilde{H}_{N}t)T_{N}^{+}\qquad
(t=1)\qquad \qquad \qquad \qquad \ \ (15)$ \newline
where the diagonal operator $\widetilde{H}_{N}$ is a member of the
longitudinal magnetization and spin order operator subspace:

$\widetilde{H}_{N}=\Omega _{0}+\stackunder{k=1}{\stackrel{N}{\sum }}\Omega
_{k}^{^{\prime }}I_{kz}+\stackrel{N}{\stackunder{l>k=1}{\sum }}%
J_{kl}^{^{\prime }}2I_{kz}I_{lz}+\stackrel{N}{\stackunder{m>l>k=1}{\sum }}%
J_{klm}^{^{\prime }}4I_{kz}I_{lz}I_{mz}+...,\ \ \ (16)$ \newline
here unit operator $E$ is omitted. All the parameters in Eqs.(14)-(16) are
determined directly from the elements $\{u_{ij}\}$ of the matrix $u(2)$
(Miao, 2000a).

In particular, for the two- and three-qubit gates (Barenco, 1995 \& Deutsch,
1989) the diagonal unitary operators can be respectively written as

$\qquad \qquad \exp (-i\widetilde{H}_{2}t)=\exp (-i\Omega _{0})\exp
(-i\Omega _{1}^{^{\prime }}I_{1z})\exp (-i\Omega _{2}^{^{\prime }}I_{2z})$

$\qquad \qquad \qquad \qquad \qquad \quad \ \ \times \exp
(-iJ_{12}^{^{\prime }}2I_{1z}I_{2z})\qquad \qquad \qquad \qquad \qquad (17)$ 
\newline
and

$\exp (-i\widetilde{H}_{3}t)=\exp (-i\Omega _{0})\stackrel{3}{\stackunder{k=1%
}{\prod }}\exp (-i\Omega _{k}^{^{\prime }}I_{kz})\stackrel{3}{\stackunder{%
l>k=1}{\prod }}\exp (-iJ_{kl}^{^{\prime }}2I_{kz}I_{lz})$

$\qquad \qquad \qquad \qquad \qquad \times \exp (-iJ_{123}^{^{\prime
}}4I_{1z}I_{2z}I_{3z})\qquad \qquad \qquad \ \quad \ \ \qquad (18)$ \newline
By using Eq.(10) the last three-body elementary propagator on the right-hand
side of Eq.(18) can be further decomposed as a sequence of six one-body
elementary propagators and three two-body diagonal elementary propagators
and hence the Deutsch$^{^{\prime }}$s three-qubit universal gate is exactly
decomposed as a sequence of six two-qubit diagonal gates $R_{kl}(\lambda
_{kl})$ and thirteen one-qubit gates and one constant phase factor.\newline
\newline
\textbf{4.2 The Grover}$^{^{\prime }}$s\textbf{\ quantum search algorithm. }

In most quantum algorithms the first step is the creation of a
superposition. This may be achieved by applying Walsh-Hadamard transform on
the groundstate in a quantum system. Any n-qubit Walsh-Hadamard transform is
constructed by the direct product of n single-qubit M matrices (Grover,
1997) and can be expressed as a sequence of one-body elementary propagators
and a constant phase factor:

$W=M_{1}\bigotimes M_{2}\bigotimes ...\bigotimes M_{n}$

$\ \ \ =\exp (i\frac{n\pi }{2})\exp (-i\pi \stackrel{n}{\stackunder{k=1}{%
\sum }}I_{kx})\exp (-i\frac{\pi }{2}\stackrel{n}{\stackunder{k=1}{\sum }}%
I_{ky})$ \newline
In addition to the Walsh-Hadamard transform the basic unitary operations
required by the Grover$^{^{\prime }}$s algorithm (Grover, 1997) are the
conditional phase shift operations represented respectively by the diagonal
unitary matrix C and R:

$\quad \ \ C_{ij}=0,\ if\ \ i\neq j;\ C_{ii}=-1;\ if\ \ i=s;\ \ C_{ii}=1,\
if\ \ i\neq s$ \newline
and $\quad R_{ij}=0,\ \ if\ \ i\neq j;\ \ R_{ii}=-1,\ \ if\ \ i=1;\ \
R_{ii}=-1,\ if\ \ i\neq 1$ \newline
The unitary operation $C$ can be further expressed in the form of
exponential operator

$C=\exp (-iH_{c}t)\qquad (t=1)$ \newline
where the representation matrix elements of the diagonal operator $H_{c}$
can be derived from the matrix $C$ as

$(H_{c})_{ij}=0,\ \ if\ \ i\neq j;\ \ (H_{c})_{ii}=\pi ,\ \ if\ \ i=s;\ \
(H_{c})_{ii}=0,\ \ if\ \ i\neq s$ \newline
Therefore, the operator $H_{c}$ is a member of the longitudinal
magnetization and spin order operator subspace and can be expressed as a sum
of the base operators of the subspace. It takes the same form as the
diagonal operator $\widetilde{H}_{N}$ of Eq.(16) but with different
parameters determined from the matrix elements $\{(H_{c})_{ij}\}$. As a
result of Eq.(11), the unitary operation $C$ can be decomposed into a
product of elementary propagators built up with the base operators of the
subspace

$\qquad C=\exp (-i\Omega _{0})\stackunder{k=1}{\prod }\exp (-i\Omega
_{k}^{^{\prime }}I_{kz})\stackunder{l>k=1}{\prod }\exp (-iJ_{kl}^{^{\prime
}}2I_{kz}I_{lz})$

$\qquad \qquad \qquad \times \stackunder{m>l>k=1}{\prod }\exp
(-iJ_{klm}^{^{\prime }}4I_{kz}I_{lz}I_{mz})...$ \newline
According to Eq.(10) the unitary operation $C$ can be further decomposed
completely as a sequence of one-body elementary propagators and the two-body
diagonal elementary propagators. The diagonal phase rotation operation $R$
can be decomposed completely in an analogous way as the unitary operation $C$%
. Thus, each of the basic unitary operations $W$, $C$, and $R$ (the
diffusion transform $D=WRW$) in any n-qubit Grover$^{^{\prime }}$s algorithm
is expressed explicitly as a sequence of one-qubit gates and the two-qubit
diagonal gates $R_{kl}(\lambda _{kl})$. This result may be helpful to
implement experimentally the algorithm with any qubits in an accessible
two-state quantum system. \newline
\newline
\textbf{4.3 The Deutsch-Jozsa}$^{^{\prime }}$\textbf{s algorithm}

\smallskip To decide certainly whether a function $f:B^{n}\longrightarrow B$
is balanced or constant, it needs to run unitary transformation $U_{f}$ only
once on the superposition (Deutsch \& Jozsa, 1992; Jozsa, 1998; Cleve, et
al. 1998):

$U_{f}:\dfrac{1}{\sqrt{2^{n}}}\stackunder{x_{i}\in B^{n}}{\sum }%
|x_{i}\rangle [\dfrac{1}{\sqrt{2}}(|0\rangle -|1\rangle )]\longrightarrow $

$\qquad \qquad \qquad \qquad \qquad \qquad \dfrac{1}{\sqrt{2^{n}}}%
\stackunder{x_{i}\in B^{n}}{\sum }(-1)^{f(x_{i})}|x_{i}\rangle [\dfrac{1}{%
\sqrt{2}}(|0\rangle -|1\rangle )]$ \newline
Therefore, $U_{f}$ has a diagonal unitary representation and can be
expressed in the form of exponential operator:

$U_{f}=\exp (-iH_{f}t)\qquad (t=1)$ \newline
where the diagonal operator $H_{f}$ has the representation matrix elements:

$(H_{f})_{ij}=0,\ \ if\ \ i\neq j;\ \ (H_{f})_{ij}=\{ 
\begin{array}{l}
0,\ \ if\ \ f(x_{i})=0 \\ 
\pi ,\ \ if\ \ f(x_{i})=1
\end{array}
$ \newline
Obviously, any N-qubit unitary operation $U_{f}$ can be readily decomposed
completely in a similar way to the unitary operation $C$ in the Grover$%
^{^{\prime }}$s algorithm.\newline
\newline
\textbf{5. Discussion}

It is widely believed that the three-qubit universal gates are sufficient to
build any quantum computation (Deutsch, 1989), but several investigators
(DiVincenzo, 1995a; Barenco, 1995; Sleator and Weinfurter, 1995) showed that
any three-qubit universal gates can be further expressed as a sequence of
the two-qubit universal gates and hence the latters are more basic units in
quantum computation. In the paper it is shown that any three- and two-qubit
universal gates as well as the XOR gates can be expressed as a simple
sequence of one-qubit gates and the two-qubit diagonal gates. These simple
gates can be constructed with natural one- and two-body interactions such as
neighbor interaction and could be readily realized experimentally in a
two-state quantum system. Therefore, the two-qubit diagonal gates should be
also proper elementary building blocks to construct conveniently any quantum
computation physically.

The decomposition for any unitary transformation of the quantum network of a
given quantum algorithm into a sequence of one-qubit gates and the two-qubit
diagonal gates provides a good scheme for the quantum algorithm to be
programmed on a quantum computer. The effective Hamiltonian of a given
quantum algorithm characterizes generally the specific properties of the
quantum algorithm and the operator algebra structure of the effective
Hamiltonian may decide how the decomposition is implemented conveniently.
Therefore, the decomposition may be achieved conveniently with the help of
the operator algebra structure of the effective Hamiltonian and the
properties of the Liouville operator space and its three operator algebra
subspaces. The explicit decomposition for any unitary transformation of
quantum computational networks into a sequence of one-qubit gates and the
two-qubit diagonal gates in an exact and unified form will be helpful for
implementing generally any N-qubit quantum computation in feasible quantum
systems and determining conveniently the time evolution of these systems in
course of quantum computing.

The effective Hamiltonian of a given quantum algorithm may also characterize
generally the complexity of a quantum algorithm. Provided that the effective
Hamiltonian consists of local interations of a quantum system subjected to
the quantum algorithm, there is certainly a quantum computational network
that can simulate efficiently the quantum computation (Lloyd, 1996). If a
classical algorithm, which may not be efficient, is designed to solve an
NP-problem in a classical digital computer and it can be translated into a
quantum algorithm by replacing irreversible logic gates with the
corresponding reversible gates according to the Bennett$^{^{\prime }}$s
suggestion (Bennett, 1973), now one wants to ask: can the NP-problem be
solved efficiently with the quantum algorithm on a quantum computer?
Evidently, this is impossible (Deutsch, 1985 \& 1989). Is there other
quantum algorithm to solve efficiently the same NP-problem? If a quantum
computational network is designed according to the mathematical structure
and characteristic of the NP-problem and the quantum mechanical laws and if
the effective Hamiltonian of the quantum network of the quantum algorithm is
local, the network can solve efficiently the NP-problem. \newline
\newline
\textbf{Acknowledgment }

This work was supported by the Hong Kong University of Science and
Technology and Professor Dr.T.Y.Tsong when author visited his research group
in the Department of Biochemistry of the Hong Kong University of Science and
Technology. \newline
\newline
\textbf{Added note}: The initial version of the paper was submitted to the
journal of Phys.Rev.Lett. on 2 February 1999 (Ref. number: LP7405 and the
receipt date: 16 February 1999). The present paper is the modified version.%
\newline
\newline
\textbf{References}\newline
Barenco, A. 1995 Proc.R.Soc.Lond. A \textbf{449}, 679 \newline
Barenco, A., Deutsch, D., Ekert, A. \& Jozsa, R. 1995a Phys.Rev.Lett.\textbf{%
\ 74}, 4083 \newline
Barenco,\ A. Bennett, C.H., Cleve, R., DiVincenzo, D.P., Margolus, N., Shor,
P.W., Sleator, T., Smolin, J.A. \& Weinfurter, H. 1995b Phys.Rev. A \textbf{%
52}, 3457 \newline
Bennett, C.H. 1973 IBM J.Res.Develop. \textbf{17}, 525 \newline
Benioff, P. 1980 J.Stat.Phys. \textbf{22}, 563 \newline
Chuang, I.L., Vandersypen, L.M.K., Zhou, X., Leung, D.W. \& Lloyd, S. 1998a
Nature \textbf{393}, 143\newline
Chuang, I.L., Gershenfeld, N.A. \& Kubinec, M. 1998b Phys.Rev.Lett. \textbf{%
80}, 3408\newline
Cirac, J.I. \& Zoller, P. 1995 Phys.Rev.Lett. \textbf{74}, 4091\newline
Cory, D.G., Fahmy, A.F. \& Havel, T.F. 1997 Proc.Natl.Acad.Sci. USA \textbf{%
94}, 1634 \newline
Cleve, R., Ekert, A., Macchiavello, C. \& Mosca, M., 1998 Proc.R.Soc.Lond. A 
\textbf{454}, 339 \newline
Deutsch, D. 1985 Proc.R.Soc.Lond. A \textbf{400}, 97 \newline
Deutsch, D. 1989 Proc.R.Soc.Lond. A \textbf{425}, 73 \newline
Deutsch, D. \& Jozsa, R. 1992 Proc.R.Soc.Lond. A \textbf{439}, 553 \newline
Deutsch, D., Barenco, A. \& Ekert, A. 1995 Proc.R.Soc.Lond. A \textbf{449},
669 \newline
DiVincenzo, D.P. 1995a Phys.Rev. A \textbf{51}, 1015 \newline
DiVincenzo, D.P. 1995b Science \textbf{270}, 255\newline
Ernst, R.R., Bodenhausen, G. \& Wokaun, A. 1987 \textit{Principles of
Nuclear }

\textit{Magnetic Resonance in\ One and Two Dimensions}.

Oxford: Oxford University Press\newline
Feynman, R.P. 1982 Int.J.Theor.Phys. \textbf{21}, 467\newline
Fredkin, E. \& Toffoli, T. 1982 Int.J.Theor.Phys. \textbf{21}, 219 \newline
Freeman, R. 1997 \textit{Spin Choreography. }Oxford: Spektrum\newline
Grover, L.K. 1997 Phys.Rev.Lett. \textbf{79}, 325\newline
Gershenfeld, N.A. \& Chuang, I.L. 1997 Science \textbf{275}, 350\newline
Jones, J.A., Mosca, M. \& Hansen, R.H. 1998 Nature \textbf{393}, 344\newline
Jozsa, R. 1998 Proc.R.Soc.Lond. A \textbf{454}, 323\newline
Kane, B.E. 1998 Nature \textbf{393}, 133 \newline
Lloyd, S. 1995 Phys.Rev.Lett. \textbf{75}, 346 \newline
Lloyd, S. 1996 Science \textbf{273}, 1073\newline
Loss, D. \& DiVincenzo, D.P. 1998 Phys.Rev. A \textbf{57}, 120\newline
Miao, X., Han, X., \& Hu, J. 1993 Sci.China A \textbf{36}, 1199\newline
Miao, X. \& Ye, C. 1997 Mol.Phys. \textbf{90}, 499 \newline
Miao, X. 2000a Molec.Phys. (in press)\newline
Miao, X. 2000b http://xxx.lanl.gov/abs/quant-ph/0003113 \newline
Makhlin, Yu., Sch$\ddot{o}$n, G. \& Shnirman, A. 1999 Nature \textbf{398},
305\newline
S$\phi $rensen, O.W., Eich, G.W., Levitt, M.H., Bodenhausen, G., \& Ernst,
R.R.

1983 Prog. NMR Spectrosc. \textbf{16}, 163 \newline
Shor, P.W. 1994 \textit{Proc.35th Ann.Symp.on Found.of Computer Science},

IEEE Comp.Soc.Press, Los Alamitos, CA, pp.124\newline
Sleator, T. \& Weinfurther, H. 1995 Phys.Rev.Lett. \textbf{74}, 4087\newline
Toffoli, T. 1981 Math.System theory \textbf{14}, 13\newline
\newline
\textbf{Appendix A}\newline
\textbf{The multiple-quantum operator algebra spaces }

A $p$-quantum operator $Q_{p}$ is defined by (Miao, et al. 1993, Miao, 2000a)

$\qquad \qquad I_{z}Q_{p}|\Psi _{r}\rangle =(M_{r}+P)Q_{p}|\Psi _{r}\rangle
\qquad (\hslash =1)\qquad \qquad \qquad \quad \quad (A1)$ \newline
where the wavefunction $|\Psi _{r}\rangle $ is an arbitrary eigenstate of
the z-component $I_{z}$ of the total spin angular momentum operator of a
spin system with its own eigenvalue $M_{r}$

$\qquad \qquad \qquad \qquad I_{z}|\Psi _{r}\rangle =M_{r}|\Psi _{r}\rangle
\qquad \qquad \qquad \qquad \qquad \qquad \qquad \ \ (A2)$ \newline
The operator $I_{z}$ is also called the total magnetic quantum operator or
the total longitudinal magnetization operator of the system. The definition
of Eq.(A1) of a $p$-quantum operator is general and independent of energy
eigenstates of the system, although the wavefunction $|\Psi _{r}\rangle $ is
also an eigenfunction of spin Hamiltonian of the system when the
contribution of Zeeman interaction to the spin Hamiltonian is dominating.
However, when spin Hamiltonian of a spin system contains non-secular
interactions the operator $I_{z}$ usually does not commute with the spin
Hamiltonian and in this case $|\Psi _{r}\rangle $ is not an eigenfunction of
the spin Hamiltonian.

The $p$-quantum operator $Q_{p}$ has an explicit physical meaning that a new
state $Q_{p}|\Psi _{r}\rangle $ generated by $Q_{p}$ acting on an arbitrary
eigenstate $|\Psi _{r}\rangle $ is also an eigenstate of the total magnetic
quantum operator $I_{z}$ and its total magnetic quantum number raises $p$
from the original one $M_{r}$. It proves easily from the definition of the $%
p $-quantum operator that the complete set of the $p$-quantum operators can
construct a \textit{linear subspace} of the Liouville operator space of the
spin system since the sum of any two $p$-quantum operators is also a $p$%
-quantum operator. In particular, the complete set of zero-quantum operators
is an operator algebra subspace of the Liouville operator space. This can be
proven simply below. By expanding the eigenstates $|\Psi _{r}\rangle $ and $%
Q_{p}|\Psi _{r}\rangle $ in terms of the complete orthogonal and normalized
eigenbase $\{|k\rangle \}$ of the operator $I_{z}$ with eigenvalues $%
\{M_{k}\}$, respectively

$\qquad \qquad |\Psi _{r}\rangle =\stackunder{k}{\sum }B_{rk}(0)|k\rangle ,$
for all $k$ with $M_{k}=M_{r}\qquad \qquad \qquad \ \ (A3)$ \newline
and

$\qquad \qquad \qquad Q_{p}|\Psi _{r}\rangle =\stackunder{k,l}{\sum }%
B_{rk}(0)C_{kl}(p)|l\rangle ,\qquad \qquad \qquad \qquad \qquad \quad \ (A4)$
\newline
where sums run over all indexes $l$ with $M_{l}=(M_{r}+P)$ and $k$ with $%
M_{k}=M_{r}$, respectively, one can prove easily that the product operator
of any two zero-quantum operators $Q_{0\alpha \newline
}$ and $Q_{0\beta }$ is still a zero-quantum operator. It follows from
Eq.(A4) that

$\qquad \qquad Q_{0\beta }Q_{0\alpha }|\Psi _{r}\rangle =\stackunder{k,l,m}{%
\sum }B_{rk}(0)C_{kl}^{\alpha }(0)C_{lm}^{\beta }(0)|m\rangle \qquad \qquad
\qquad \quad \ \ (A5)$ \newline
where sums run over all indexes $k,l,m$ with $M_{k},M_{l},M_{m}=M_{r}$.
Evidently, the product operator $Q_{0\beta }Q_{0\alpha }$ is still a
zero-quantum operator since all the eigenstates of the operator $I_{z}$ on
the right-hand side of Eq.(A5) have the same eigenvalue equal to $M_{r}$ of $%
|\Psi _{r}\rangle ,$

$\qquad \qquad I_{z}(Q_{0\beta }Q_{0\alpha })|\Psi _{r}\rangle
=M_{r}(Q_{0\beta }Q_{0\alpha })|\Psi _{r}\rangle .$ \newline
Therefore, all the zero-quantum operators form an operator algebra subspace
of the Liouville operator space. As a direct result, the power operator $%
(Q_{0})^{n}\ $(n=1,2,...) of a zero-quantum operator $Q_{0}$ is a
zero-quantum operator and moreover, the exponential operator $\exp (\pm
i\lambda Q_{0})$ of a Hermite zero-quantum operator $Q_{0}$ is a
zero-quantum unitary operator and can be expressed as a sum of base
operators $\{Q_{0k}\}$ of the zero-quantum operator subspace

$\qquad \exp (\pm i\lambda Q_{0})=\stackunder{k}{\sum }\frac{(\pm i\lambda
)^{n}}{n!}(Q_{0})^{n}=\stackunder{k}{\sum }f_{k}(\pm \lambda )Q_{0k}\qquad
\qquad \qquad \qquad (A6)$ \newline
where $f_{k}(\pm \lambda )$ are coefficients. There is an important property
of the zero-quantum operator that any p-quantum operator does not change its
quantum coherence order when it is acted on by an arbitrary zero-quantum
operator. The proof for the property is simple. According to the definition
Eq.(A1) of a $p$-quantum operator and Eqs.(A2)-(A4), one has

$\quad \quad Q_{0\beta }Q_{p}Q_{0\alpha }|\Psi _{r}\rangle =\stackunder{%
k,l,m,n}{\sum }B_{rk}(0)C_{kl}^{\alpha }(0)C_{lm}(p)C_{mn}^{\beta
}(0)|n\rangle \qquad \qquad \ \ \ (A7)$ \newline
where sums run over indexes $k$ and $l$ with $M_{k}=M_{l}=M_{r}$ as well as $%
m$ and $n$ with $M_{m}=M_{n}=$ $(M_{r}+P)$, respectively. Because all the
eigenstates $|n\rangle $ on the right-hand side of Eq.(A7) have the same
eigenvalue $(M_{r}+P),$ the state $Q_{0\beta }Q_{p}Q_{0\alpha }|\Psi
_{r}\rangle $ is an eigenstate of the operator $I_{z}$ and its own
eigenvalue equals $(M_{r}+P)$,

$\qquad \qquad I_{z}(Q_{0\beta }Q_{p}Q_{0\alpha })|\Psi _{r}\rangle
=(M_{r}+P)(Q_{0\beta }Q_{p}Q_{0\alpha })|\Psi _{r}\rangle $ .\newline
Therefore, the product operator $Q_{0\beta }Q_{p}Q_{0\alpha }$ is a $p$%
-quantum operator, indicating that any p-quantum operator keeps its quantum
coherence order unchanged when it is acted on by a zero-quantum operator.
Particularly, any zero-quantum operator can be transferred into a sum of the
base operators of the zero-quantum operator subspace by making a
zero-quantum unitary transformation. This is really a direct consequence of
the closed property of the zero-quantum operator subspace.

In particular, it follows from the definition of Eq.(A1) of a zero-quantum
operator that all the zero-quantum operators that are commutable with each
other and also commute with the total magnetic quantum operator $I_{z}$
should form an operator algebra subspace of the zero-quantum operator
subspace. This subspace is called the longitudinal magnetization and spin
order operator subspace.

An even-order multiple-quantum operator $Q_{ek}$ is defined by

$\qquad \qquad \ \ \ Q_{ek}=\stackunder{p}{\sum }B_{kp}Q_{2p}\qquad (p=0,\pm
1,\pm 2,...)\qquad \qquad \qquad \qquad (A8)$ \newline
where $B_{kp}$ are coefficients and the operators $Q_{2p}$ are $2p$-quantum
operators:

$\qquad \qquad \qquad I_{z}Q_{2p}|\Psi _{r}\rangle =(M_{r}+2P)Q_{2p}|\Psi
_{r}\rangle \qquad \qquad \qquad \qquad \qquad (A9)$\newline
The definition (A8) of an even-order multiple-quantum operator shows that
the complete set of the even-order multiple-quantum operators is \textit{a
linear subspace} of the Liouville operator space. Here will prove further
that all the even-order multiple-quantum operators form an operator algebra
subspace of the Liouville operator space. First of all, the product operator
of any two $2p$-quantum operators is an even-order multiple-quantum
operator. It can be found easily from Eqs.(A1)-(A4) that

$\qquad \qquad I_{z}Q_{2q}Q_{2p}|\Psi _{r}\rangle
=[M_{r}+2(p+q)]Q_{2q}Q_{2p}|\Psi _{r}\rangle \qquad \qquad \qquad \ \ (A10)$ 
\newline
This equation indicates that the product operator $Q_{2q}Q_{2p}$ is a $%
2(p+q) $-quantum operator, i.e., an even-order multiple-quantum operator. It
turns out from Eq.(A10) and the definition Eq.(A8) of an even-order
multiple-quantum operator that the product operator $Q_{ek}Q_{el}$ of any
two even-order multiple-quantum operators $Q_{ek}$ and $Q_{el}$ is still an
even-order multiple- quantum operator. Therefore, all the even-order
multiple-quantum operators can form an operator algebra subspace of the
Liouville operator space.

Obviously, it follows from the definition Eq.(A8) of an even-order
multiple-quantum operator that the even-order multiple-quantum operator
subspace contains the whole zero-quantum operator subspace.

There are some important properties of the even-order multiple-quantum
operator subspace. One of which is that the exponential operator $\exp (\pm
i\lambda Q_{e})$ constructed with a Hermite even-order multiple-quantum
operator $Q_{e}$ is still an even-order multiple-quantum unitary operator
and can be expressed as a sum of base operators $\{Q_{ek}\}$ of the operator
subspace

$\qquad \exp (\pm i\lambda Q_{e})=\stackunder{p}{\sum }f_{p}(\pm \lambda
)Q_{ep}.\qquad (p=0,\pm 1,\pm 2,...)$ \newline
where $f_{p}(\pm \lambda )$ are coefficients. Another is that any even-order
multiple-quantum operator is transferred into a sum of base operators of the
operator subspace when it is acted on by an even-order multiple-quantum
operator. These properties are obviously a direct consequence of the closed
property of the even-order multiple-quantum operator algebra subspace.

The properties of the longitudinal magnetization and spin order, the
zero-quantum, and the even-order multiple-quantum operator subspace of the
Liouville operator space of a two-state quantum system like a spin system
may be helpful for simplifying the decomposition of the time-evolutional
propagator and the determination of unitary time evolution of the quantum
system, and the decomposition of unitary transformations of qauntum
computation into a sequence of one-qubit gates and the two-qubit diagonal
quantum gates.\newline
\newline
\textbf{Appendx B}

In a cold trapped ion system the two-qubit diagonal quantum gate\newline
$R_{mn}(\lambda _{mn})=\exp (-i\lambda _{mn}2I_{mz}I_{nz})$ may be
constructed by six laser pulses. This elementary gate is really prepared
with indirect interaction between a pair ions in the system since the
interaction is set up by an intermediate media, i.e., phonon, while the
direct interaction occurs between ions and phonons in the system. Adopting
Cirac and Zoller$^{^{\prime }}$s notation (Cirac \& Zoller,1995), the
elementary gate for a pair of ions $m$ and $n$ can be explicitly prepared by
the following laser pulse sequence:

$R_{mn}(\lambda _{mn})=U_{m}^{1,0}(\phi _{3})V_{n}^{1}(\phi
_{2})U_{n}^{1,1}(\theta _{2})U_{n}^{1,1}(\theta _{1})V_{n}^{1}(\phi
_{1})U_{m}^{1,0}(\phi _{0})$ \newline
where the parameters $\phi _{i}\ (i=1,2)$ and $\theta _{i}\ (i=1,2)$ are the
laser phases applied to the ion $n$ and $\phi _{i}\ (i=0,3)$ are the ones
applied to the ion $m,$ and they are not independent but subjected to the
following relations:

$\phi _{0}-\phi _{3}=\pi +2(\phi _{1}-\phi _{2}),$ $\quad \theta _{1}-\theta
_{2}=\pi +4(\phi _{1}-\phi _{2}).$\newline
The parameter $\lambda _{mn}$ can be determined by

$\lambda _{mn}=2\pi -2(\phi _{1}-\phi _{2})$\newline
Therefore, by adjusting suitably the phase difference $(\phi _{1}-\phi _{2})$
of the lasers applied to the ion $n$ the desired parameter $\lambda _{mn}$
can be obtained.\newline
\newline
\newline
\textbf{Figure 1.} The preparation for the direct two-body elementary
propagator: $SE_{n}=R_{kl}(\lambda _{kl})=\exp (-i\lambda
_{kl}2I_{kz}I_{lz})\ (\lambda _{kl}=4^{n}\pi J_{kl}\tau )$ in a coupled
N-spin (I=1/2) system with Hamiltonian of Eq.(12). The N spins except spins $%
k$ and $l$ are divided into several groups $\{p\},\ \{s\},\ ...,\ \{w\},$
where both two coupled spins $k\ $and $l$ are not coupled with those spins
of group $\{p\}$ and there may be interaction among these groups but is not
coupling between any arbitrary two spins in each of these groups. The spin
echo sequence $SE_{1}$ with selective $180{{}^{\circ }}$ RF pulses applied
simultaneously to the two spins $k$ and $l\ $and all the spins of group $%
\{p\}$ refocuses their chemical shifts (one-body interactions) and undesired
two-body interactions with any other groups but leaves the desired two-body
interaction $2I_{kz}I_{lz}$. The second sequence $SE_{2}$ with four SE$_{1}$
units and selective $180{{}^{\circ }\ }$pulses applied to spins of group $%
\{s\}$ refocuses further one- and two-body interactions of these spins with
the rest groups. Finally in sequence SE$_{n}$ all the undesired one- and
two-body terms are refocused but only the desired term $2I_{kz}I_{lz}$ is
retained and hence $R_{kl}(\lambda _{kl})$ is obtained.

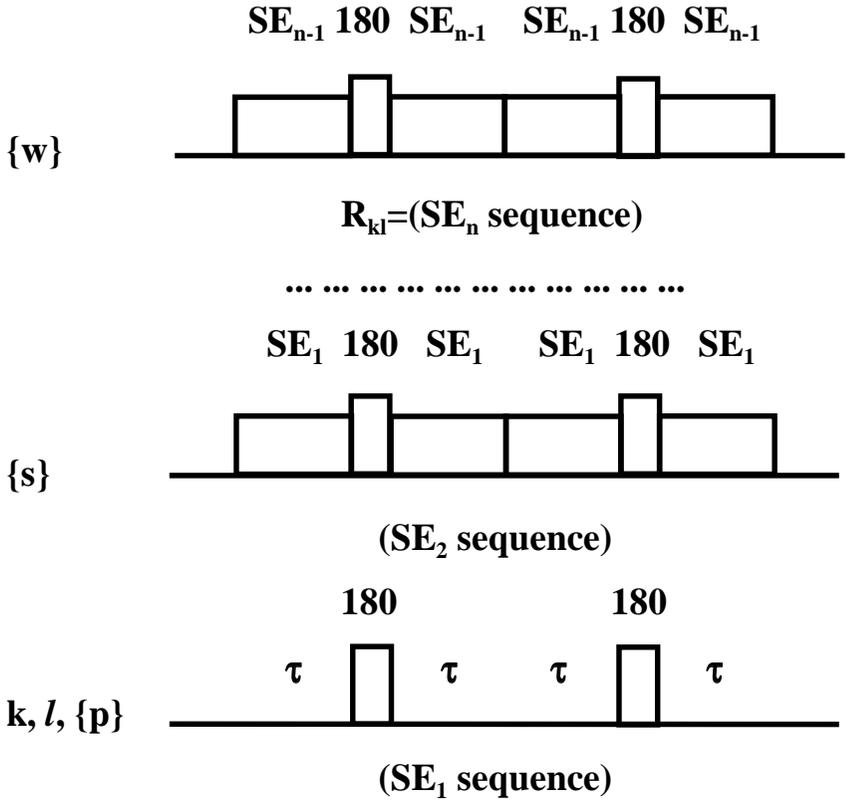
\end{document}